\documentclass[aps,superscriptaddress,pre,amsmath,amssymb]{revtex4-2}

\usepackage{graphicx,amsmath,amssymb,bm,braket,dcolumn}
\usepackage[colorlinks=true,allcolors=blue]{hyperref}

\newcommand{\new}{\nonumber\\}

\newcommand{\tv}{\tilde{v}}
\newcommand{\tb}{\tilde{\beta}}
\newcommand{\tS}{\tilde{\Sigma}}
\newcommand{\tphi}{\widetilde{\varphi}}

\newcommand{\hA}{\hat{A}}

\newcommand{\ave}[1]{\left\langle #1\right\rangle}

\newcommand{\ox}{\overline{x}}
\newcommand{\oy}{\overline{y}}

\newcommand{\omu}{\overline{\mu}}
\newcommand{\onu}{\overline{\nu}}

\begin{document}

\title{High-dimensional theory of the glass transition revisited: 
 hopping and local defects}

\author{Harukuni Ikeda}
\email{harukuni.ikeda@yukawa.kyoto-u.ac.jp}
\affiliation{Yukawa Institute for Theoretical Physics, Kyoto University,
Kyoto 606-8502, Japan 
}

\author{Francesco Zamponi}
\affiliation{Dipartimento di Fisica, Sapienza Universit\`a di Roma, P.le Aldo Moro 5, 00185 Rome, Italy}

\date{\today}

\begin{abstract}
The replicated liquid theory provides a microscopic mean-field
description of the glass transition by combining the density functional
theory of liquids with the replica method originally developed for spin
glasses.  In the conventional replica liquid theory, a glassy state is
described by assuming that particles in different replicas undergo
vibrational motion around common centers of mass, thereby forming
molecules that contain one particle from every replica. Here we revisit
this assumption by allowing each molecule to contain only a subset of
replicas. This generalized formulation describes particle-level replica
mismatches, which may be associated with non-vibrational motions such as
particle hopping. We apply the theory to high-dimensional hard and
harmonic spheres, where the mean-field description is expected to become
exact. For hard spheres, replica mismatches destabilize the glassy
metastable state and shift the dynamical transition to a significantly
higher packing fraction, while leaving the leading thermodynamic glass
transition unchanged. The revised hard-sphere dynamical transition reproduces the leading
term $d\log d/2^{d+1}$ in the packing lower bound of Campos, Jenssen,
Michelen, and Sahasrabudhe. Whether a reversible insertion--deletion
process arrests at this density remains open.  For harmonic spheres at finite temperature, the
glassy state contains a finite replica-mismatch fraction even at the
thermodynamic ideal-glass transition, thereby shifting the transition
point from that predicted by the conventional replica ansatz.
\end{abstract}

\maketitle
\tableofcontents

\section{Introduction}

The glass transition is a long-standing problem in statistical and
condensed matter physics. When a liquid is cooled or compressed while
avoiding crystallization, its relaxation time and viscosity grow
dramatically, until it eventually behaves as an amorphous solid. A
central question is whether this phenomenon can be described as the
emergence of metastable amorphous states and an associated thermodynamic
transition~\cite{Debenedetti2001,cavagna2009,berthierbiroli2011}.

The mean-field scenario of the glass transition was originally developed
in solvable spin-glass models, most notably the $p$-spin spherical
model~\cite{kirkpatrick1987,pott1987}.  In this scenario, the
free-energy landscape splits, upon cooling, into an exponentially large
number of metastable states. The emergence of these states is associated
with a dynamical
transition~\cite{leutheusser1984,Bengtzelius1984,KW87,maimbourg2016,szamel2017},
while the point at which their configurational entropy vanishes is
identified with the thermodynamic, or ideal-glass, transition. Although
finite-dimensional systems may avoid a sharp dynamical singularity
through activated processes, this mean-field picture remains an
essential reference point for understanding glassy dynamics and
thermodynamics~\cite{Adam1965,scaling1989,Bouchaud2004}.

The replica liquid theory provides a microscopic realization of this
scenario for structural glasses by combining the replica method
developed for spin-glasses and the density functional theory of
liquids~\cite{SSW85,KW87,KT88,Mezard1986,simpleliquid,monasson1995,Mzard1996,mzard1999,PZ2010,simpleglass}.
One introduces replicas of the liquid and describes a metastable glass
state by assuming that particles in different replicas are localized
around common centers of mass, forming molecules made of replicated
particles. The replicated system is then treated as a liquid of these
molecules. This theory has been particularly successful for
high-dimensional spheres, where the virial expansion becomes controlled
and the mean-field description is expected to become
exact~\cite{KW87,PZ2010,simpleglass,Kurchan2012,Kurchan2013,Charbonneau2014,Sellitto2013,Altieri2018,Ikeda2021,kim2025}.

In the conventional replica liquid theory, each molecule contains one
particle from every replica, and the replicated particles are localized
around a common center~\cite{mzard1999,PZ2010,simpleglass}.  This is
similar to an Einstein crystal where each atom in the glass is
constrained to be localized around its reference position.  It therefore
excludes particle-level mismatches, where some particles do not form
replicated molecules with other replicas. Here we revisit the theory by
allowing each molecule to contain only a subset of replicas, leading to
fluctuations of molecular sizes. The conventional theory is recovered
when all molecules contain all replicas.

There are two main motivations for introducing this extension. The first
comes from high-dimensional sphere packing. Recent mathematical progress
has established a rigorous lower bound for random sphere packings whose
density is significantly higher than the dynamical transition predicted
by the conventional replica liquid theory~\cite{campos2023new}. This
bound lies a factor of two below the `Glass Close Packing' density $d\log d\,2^{-d}$, 
corresponding to the highest density that can be reached by 
adiabatic compression of the amorphous liquid state~\cite{PZ06a,PZ2010}, 
a comparison already noted by Campos {\it et al.}~\cite{campos2023new}.
Here, we identify the scale of Campos {\it et al.} with the dynamical transition 
associated with the stability of the inter-replica overlap.
Note that Klartag has since proven a lower bound $c\,d^2 2^{-d}$, with a
universal constant $c>0$, for lattice packings in every dimension $d\geq
2$~\cite{Klartag2026}. This bound asymptotically exceeds both glass
density scales, which are replica predictions for amorphous states
rather than bounds on the densest packing; the result of Campos {\it et
al.} remains relevant here because it is obtained from a random
construction.  The second motivation
comes from numerical simulations of glass forming models, where
localized excitations and hopping-like rearrangements have been observed
even inside the ideal-glass
phase~\cite{ozawa2018,biroli2021interplay,jung2025numerical}.  This
suggests that an ideal glass should not necessarily be represented by a
purely vibrational state in which all replicated particles remain
localized around common molecular centers.

In the following, we call ``conventional replica theory'' that in which
each molecule contains one atom of each replica, and ``dissociation
replica theory'' that in which molecules can dissociate.  We formulate
the dissociation theory and apply it to high-dimensional hard and
harmonic spheres. For hard spheres, molecular-size fluctuations
destabilize the conventional metastable glass solution and shift the
dynamical transition to a significantly higher packing fraction. The
shifted transition density agrees, at leading order in high dimensions,
with the recent lower bound for sphere packings obtained by
Campos {\it et al.}~\cite{campos2023new}: our hard-sphere spinodal
reproduces the leading term $d\log d/2^{d+1}$ in the packing lower
bound of Theorem~1.1 of Ref.~\cite{campos2023new}. Whether a reversible
insertion--deletion process (grandcanonical dynamics) arrests at this density remains open, although we believe that 
a dynamical treatment similar to that of Ref.~\cite{maimbourg2016} would show that such a result holds.
A non-rigorous estimate supporting this belief, based on the competition
between the refilling of a vacated cage and the insertion of a particle
elsewhere, is given in Appendix~\ref{app:dyn_heuristic}.
For harmonic spheres at finite temperature, the
dissociation theory predicts that the glassy solution contains a finite
replica-mismatch fraction. This fraction becomes exponentially small at
low temperature, but remains finite at the thermodynamic ideal-glass
transition point within the present theory. Consequently, the
conventional full-replica molecular ansatz overestimates the
configurational entropy.

The rest of the paper is organized as follows. In
Section~\ref{sec:theory} we first give a schematic overview of the
molecular dissociation construction, and we then formulate the replica
liquid theory with molecular dissociation and derive the corresponding
free-energy functional. In Sec.~\ref{sec:results}, we apply the theory
to high-dimensional hard and harmonic spheres, and analyze the dynamical
and thermodynamic glass transitions.  Section~\ref{sec:summary}
summarizes the results and discusses their implications. Technical
details and connections to independent-set problems on random graphs are
given in the Appendices.

\section{Theory}
\label{sec:theory}

\subsection{Sketch of the framework}
\label{sketch}

\begin{figure}[t]
\centering \includegraphics[width=10cm]{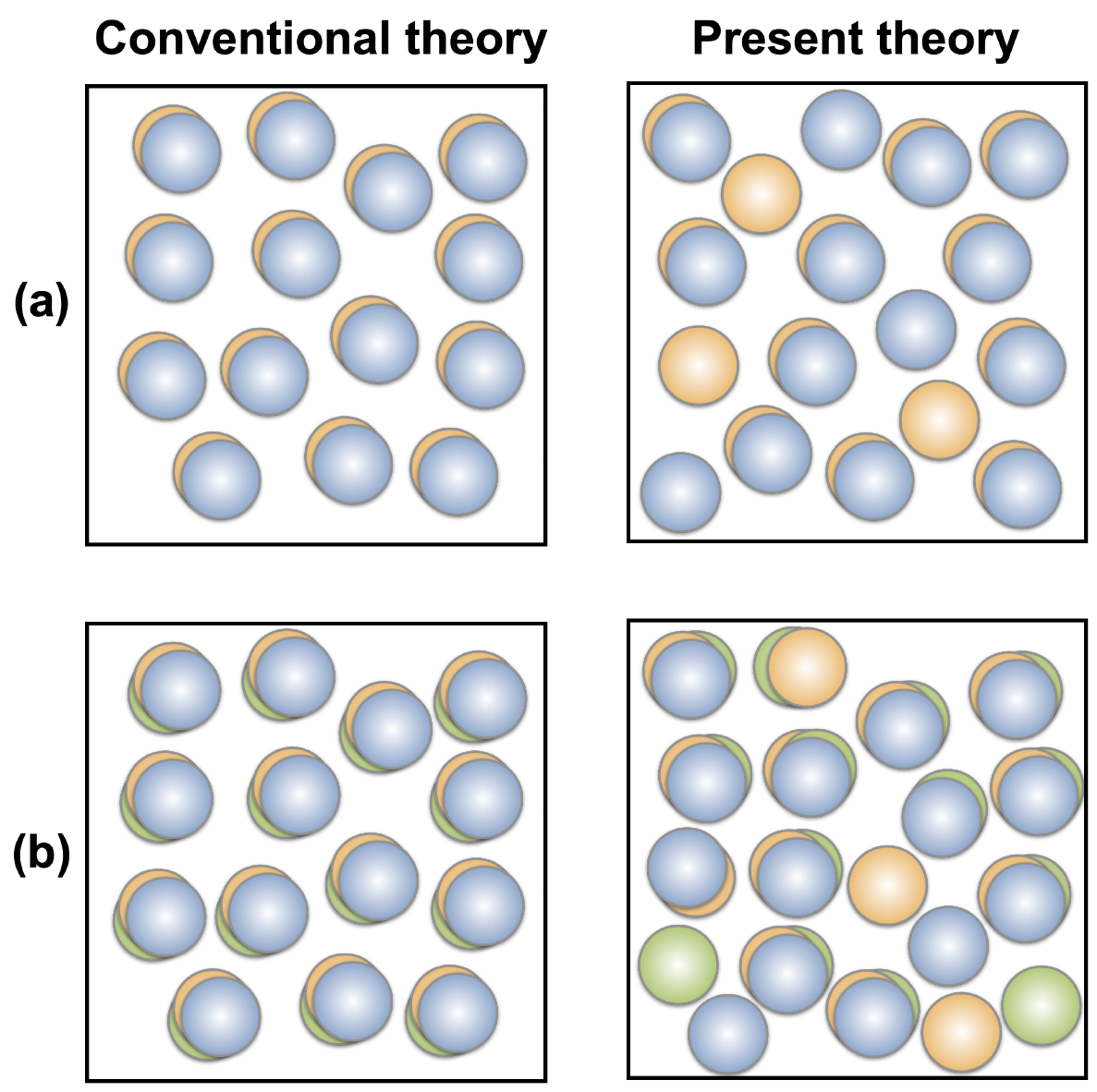} \caption{ Schematic
illustration of the conventional replica liquid theory and of the
molecular dissociation construction introduced in this work. Panel (a)
shows the case $m=2$, and panel (b) the case $m=3$. Disks of different
colors denote particles in different replicas. In the conventional
theory (left panels), every molecule contains one particle from each
replica. In the present theory (right panels), a molecule may contain
any non-empty subset of replicas. Thus, for $m=2$, there are $2^2-1=3$
possible molecular types, $(\mu^1,\mu^2)=(1,1),(1,0),(0,1)$, while for
$m=3$ there are $2^3-1=7$ possible molecular types,
$(\mu^1,\mu^2,\mu^3)=(1,1,1),(1,1,0),(1,0,1),(0,1,1),(1,0,0),(0,1,0),
(0,0,1)$.  } \label{fig:sketch}
\end{figure}

We first give a schematic overview of the framework before presenting
the formal derivation. Figure~\ref{fig:sketch} illustrates the molecular
construction in the conventional replica liquid theory and in the
present dissociation theory for $m=2$ and $m=3$ replicas.

In the conventional replica liquid theory, a glassy metastable state is
described by replicated molecules. Each molecule contains one particle
from every replica, and the particles in the molecule are localized
around a common molecular center. Thus, for $m=2$, every molecule is a
two-replica pair, while for $m=3$, every molecule is a three-replica
triplet, as shown in the left panels of Fig.~\ref{fig:sketch}.  Because
within replica theory distinct replicas physically correspond to pairs
of independent configurations belonging to the same metastable
state~\cite{monasson1995}, this ansatz physically corresponds to the
assumption that within each metastable glass, an atom always remains
near its initial position, and no hopping is possible.

In the present work, we generalize this construction by allowing a
molecule to contain only a subset of replicas. For $m=2$, molecules can
therefore be either two-replica pairs or single-replica objects. For
$m=3$, molecules can be triplets, two-replica pairs, or single-replica
objects. This introduces fluctuations in the molecular size and provides
a simple way to describe particle-level mismatches between replicas
within the replica liquid theory framework, physically corresponding to
hopping processes within the metastable glass phase, which allow atoms
to hop away from the original position.  The conventional replica liquid
theory is recovered when all molecules contain all replicas.

In the following, we formulate this idea quantitatively. We introduce a
binary molecular label specifying which replicas participate in each
molecule, construct the corresponding replicated free-energy functional,
and analyze the consequences of molecular dissociation for
high-dimensional hard and harmonic spheres.

\subsection{Molecular dissociation construction}

In this section, we formulate the replica liquid theory with fluctuating
molecular sizes.  Following the standard strategy of the replica liquid
theory, we introduce $m$ replicas of each
atom~\cite{monasson1995,mzard1999,PZ2010}.  In the conventional theory,
each molecule contains one particle from every replica.  Here we relax
this constraint and allow a molecule to contain only a subset of
replicas.  We thus introduce generalized replicated molecules that may
contain only a subset of replicas. We then derive the corresponding
free-energy functional within the second-virial approximation and impose
a Gaussian molecular ansatz. Finally, we take the high-dimensional limit
for harmonic spheres and introduce a one-parameter binomial ansatz, from
which the configurational entropy is obtained.

We label the type of a molecule by a binary vector
\begin{align}
 \omu = \{\mu^1,\mu^2,\cdots,\mu^m\},
\end{align}
where $\mu^a=1$ means that the molecule contains a particle from replica
$a$, while $\mu^a=0$ means that it does not. The molecular size in
replica space is
\begin{align}
 |\omu| = \sum_{a=1}^m \mu^a .
\end{align}
The empty molecule, $|\omu|=0$, is excluded. We denote by $N(\omu)$ the
number of molecules of type $\omu$, and write
\begin{align}
 N(\omu)=N g(\omu).
\end{align}
The coefficients $g(\omu)$ are not normalized as probabilities. The
normalization condition is imposed by requiring that each replica
contains exactly $N$ particles:
\begin{align}
 N = \sum_{\omu} N(\omu)\mu^a,
 \qquad a=1,\cdots,m .
\end{align}
Equivalently,
\begin{align}
 1 = \sum_{\omu} g(\omu)\mu^a .
\end{align}

In the following, we impose replica symmetry and assume that $g(\omu)$
depends only on the molecular size $|\omu|$:
\begin{align}
 g(\omu)=g_{|\omu|}.
\end{align}
The normalization condition then becomes
\begin{align}
1 = \sum_{k=1}^m \binom{m-1}{k-1} g_k .
\label{eq:normg}
\end{align}
The conventional replica liquid theory is recovered by the fixed-size
choice $g_k=\delta_{k m}$, whereas the completely unbound liquid
corresponds to $g_k=\delta_{k1}$.

\subsection{Replicated free-energy functional}

Let
\begin{align}
 \ox = \{x^1,\cdots,x^m\},
 \qquad
 \ox_{\omu} = \{x^a\}_{a:\mu^a=1}.
\end{align}
The interaction between two molecules of types $\omu$ and $\onu$ is
defined as
\begin{align}
V_{\omu,\onu}(\ox_{\omu},\oy_{\onu})
 = \sum_{a=1}^m \mu^a\nu^a v(x^a-y^a) ,
\end{align}
where $v(x)$ is the atomic pair potential.
Thus, an interaction exists only when both molecules contain particles
belonging to the same replica. We consider the grand-canonical ensemble
of the generalized molecular system,
\begin{align}\label{eq:XiGC}
\Xi =
\sum_{N_{\rm mol}=0}^\infty \frac{1}{N_{\rm mol}!}
\left(
\prod_{i=1}^{N_{\rm mol}}
\sum_{\omu_i:|\omu_i|>0}\int d\ox_{\omu_i}^i
\right)
\exp\left[
-\beta\sum_i \phi_{\omu_i}(\ox_{\omu_i}^i)
-\beta\sum_{i<j}
V_{\omu_i,\omu_j}(\ox_{\omu_i}^i,\ox_{\omu_j}^j)
\right], 
\end{align}
where $N_{\rm mol}=\sum_{|\omu|>0}N(\omu)$ is the total number of
molecules. The auxiliary field $\phi_{\omu}(\ox_{\omu})$ is conjugate
to the density field of molecules of type $\omu$:
\begin{align}
 \rho_{\omu}(\ox_{\omu})
 =
 -\frac{1}{\beta}
 \frac{\delta \log\Xi}{\delta \phi_{\omu}(\ox_{\omu})}.
\end{align}
Performing the Legendre transform and retaining terms up to the second
virial order~\cite{simpleliquid,PZ2010,simpleglass}, we obtain
\begin{align}
-\beta F
&=
\sum_{\omu:|\omu|>0}
\int d\ox_{\omu}\,
\rho_{\omu}(\ox_{\omu})
\left[
1-\log\rho_{\omu}(\ox_{\omu})
\right]
\nonumber\\
&\quad
+\frac{1}{2}
\sum_{\omu,\onu:|\omu|,|\onu|>0}
\int d\ox_{\omu}d\oy_{\onu}\,
\rho_{\omu}(\ox_{\omu})
\rho_{\onu}(\oy_{\onu})
f_{\omu\onu}(\ox_{\omu},\oy_{\onu}),
\label{eq:f}
\end{align}
where the Mayer function is
\begin{align}
f_{\omu\onu}(\ox_{\omu},\oy_{\onu})
=
\exp\left[
-\beta V_{\omu,\onu}(\ox_{\omu},\oy_{\onu})
\right]-1 .
\end{align}
In high dimensions, higher-order virial terms are negligible for the
glass transition of hard and soft spheres~\cite{PZ2010,simpleglass}.
They could possibly be included and partially resummed to obtain
approximate finite-dimensional theories~\cite{PZ2010,MZ16}.

\subsection{Gaussian molecular ansatz}

We now impose a Gaussian ansatz for molecules of each
type~\cite{PZ2010}:
\begin{align}
 \rho_{\omu}(\ox_{\omu})
 =
 \rho g_{|\omu|}
 \int dX
 \prod_{a:\mu^a=1}
 \gamma_A(x^a-X),
 \qquad
 \gamma_A(r)
 =
 \frac{1}{(2\pi A)^{d/2}}
 \exp\left(-\frac{r^2}{2A}\right).
\label{eq:gaussian_ansatz}
\end{align}
Here $X$ is the molecular center and $A$ is the cage size. Substituting
Eq.~\eqref{eq:gaussian_ansatz} into the ideal-gas part of
Eq.~\eqref{eq:f}, we obtain the ideal contribution to the free energy
per particle:
\begin{align}
-\beta f^{\rm id}
&=
\sum_{k=1}^m \binom{m}{k}g_k
\left(
1-\log\rho-\log g_k
\right)
\nonumber\\
&-
\sum_{k=2}^m \binom{m}{k}g_k
\int \frac{d\ox_k}{V}
\rho_k(\ox_k)\log\rho_k(\ox_k),
\end{align}
where
\begin{align}
\rho_k(\ox_k)
=
\int dX
\prod_{a=1}^k \gamma_A(x^a-X).
\end{align}
The Gaussian integral gives~\cite{PZ2010}
\begin{align}
\int d\ox_k\,\rho_k(\ox_k)\log\rho_k(\ox_k)
=
V\left[
\frac{d}{2}(1-k)\log(2\pi A)
-\frac{d}{2}\log k
+\frac{d}{2}(1-k)
\right],
\end{align}
and hence
\begin{align}
-\beta f^{\rm id}
&=
\sum_{k=1}^m \binom{m}{k}g_k
\left(
1-\log\rho-\log g_k
\right) \nonumber \\
&-
\sum_{k=2}^m \binom{m}{k}g_k
\left[
\frac{d}{2}(1-k)\log(2\pi A)
-\frac{d}{2}\log k
+\frac{d}{2}(1-k)
\right].
\label{eq:fid_general}
\end{align}

We next compute the interaction term. If two molecules have no common
replica, namely if $\mu^a\nu^a=0$ for all $a$, then $f_{\omu\onu}=0$.
Thus, only molecules sharing at least one replica contribute. Consider a
molecule $\omu$ of size $k=|\omu|$. By replica symmetry, we may assume
that the first $k$ components of $\omu$ are equal to one. For the second
molecule $\onu$, let $\ell$ be the number of replicas shared with
$\omu$, and let $s$ be the number of replicas contained in $\onu$ but
not in $\omu$. Then $|\onu|=\ell+s$, with $1\leq \ell\leq k$ and $0\leq
s\leq m-k$. After integrating over the coordinates that are not shared,
the excess free energy becomes
\begin{align}
-\beta f^{\rm ex}
=
\frac{\rho}{2}
\sum_{k=1}^m \binom{m}{k}g_k
\sum_{\ell=1}^k
\sum_{s=0}^{m-k}
\binom{k}{\ell}\binom{m-k}{s}
g_{\ell+s}
\int dr\,
\left[
q_A(r)^\ell-1
\right],
\label{eq:fex_general}
\end{align}
where
\begin{align}
q_A(r)
=
\int du\,\gamma_{2A}(r-u)\exp[-\beta v(u)].
\end{align}
Equations~\eqref{eq:fid_general} and~\eqref{eq:fex_general} give the
free-energy functional for an arbitrary replica-symmetric
molecular-size distribution $\{g_1,\cdots, g_m\}$.

\subsection{High-dimensional harmonic spheres}

We now specialize the theory to high-dimensional harmonic spheres. The
pair interaction is
\begin{align}
v(r)=
\begin{cases}
\displaystyle
\frac{\kappa}{2}\left(1-\frac{r}{D}\right)^2,
& r\leq D,\\[2mm]
0,
& r>D,
\end{cases}
\end{align}
where $D$ is the particle diameter and $\kappa$ is the stiffness. The
packing fraction is
\begin{align}
\varphi
=
\rho v_d\left(\frac{D}{2}\right)^d, \qquad v_d=\frac{\pi^{d/2}}{\Gamma(d/2+1)},
\end{align}
where $v_d$ is the volume of a $d$-dimensional unit sphere.
For reasons that will become clear in the following,
we focus on the density scale
$ \varphi \sim 2^{-d}d\log d$
and introduce
\begin{align}
 \widetilde{\varphi}
 =
 \frac{\varphi}{2^{-d}d\log d},
 \qquad
 \hat A
 =
 \frac{d^2 A}{D^2},
 \qquad
 \tilde{\beta}
 =
 \frac{\beta\kappa}{d^2}.
\end{align}
At this scale,
\begin{align}
 \log\rho
 =
 \frac{d}{2}\log d+o(d\log d),
 \qquad
 d\log A
 =
 -2d\log d+o(d\log d).
\end{align}
Keeping only the leading $O(d\log d)$ terms in
Eq.~\eqref{eq:fid_general}, we find
\begin{align}
-\beta f^{\rm id}
\simeq
\frac{d\log d}{2}
\sum_{k=1}^m \binom{m}{k}g_k(1-2k)
=
\frac{d\log d}{2}
\left[
\sum_{k=1}^m \binom{m}{k}g_k
-2m
\right].
\label{eq:fid_hd}
\end{align}

The interaction term is controlled by
\begin{align}
\rho\int dr\,\left[q_A(r)^\ell-1\right].
\end{align}
At the scale of the dynamical transition of the conventional molecular
ansatz, $\varphi=O(2^{-d}d)$, this term is only $O(d)$, and is therefore
subleading compared with the $O(d\log d)$ ideal contribution in the
present generalized variational space. This motivates the higher density
scale $\varphi\sim 2^{-d}d\log d$, where the ideal and interaction terms
compete. At this density scale, the cage size vanishes asymptotically
$\hA = O(1/\log d)$~\cite{PZ2010}, and we may approximate
\begin{align}
 q_A(r)\simeq e^{-\beta v(r)}
\end{align}
at leading order. Using the standard high-dimensional change of
variables~\cite{simpleglass}
\begin{align}
 r = D\left(1+\frac{h}{d}\right),
\end{align}
we obtain
\begin{align}
\rho \int dr\,\left[q_A(r)^\ell-1\right]
&\simeq
\rho \int dr\,
\left[
e^{-\ell\beta v(r)}-1
\right]
\nonumber\\
&=
\rho \Omega_d
\int_0^\infty dr\,r^{d-1}
\left[
e^{-\ell\beta v(r)}-1
\right]
\nonumber\\
&=
\frac{\rho\Omega_dD^d}{d}
\int_{-\infty}^{\infty}dh
\left(1+\frac{h}{d}\right)^{d-1}
\left[
e^{-\ell\tilde{\beta}\tilde v(h)}-1
\right]
\nonumber\\
&\simeq
2^d\varphi
\int_{-\infty}^{0}dh\,e^h
\left[
e^{-\ell\tilde{\beta}\tilde v(h)}-1
\right],
\end{align}
where
\begin{align}
 \tilde v(h)=\frac{h^2}{2}.
\end{align}
Therefore, Eq.~\eqref{eq:fex_general} becomes
\begin{align}
-\beta f^{\rm ex}
\simeq
d\log d\,
\frac{\widetilde{\varphi}}{2}
\sum_{k=1}^m \binom{m}{k}g_k
\sum_{\ell=1}^k
\sum_{s=0}^{m-k}
\binom{k}{\ell}\binom{m-k}{s}
g_{\ell+s}
\int_{-\infty}^0 dh\,e^h
\left[
e^{-\ell\tilde{\beta}\tilde v(h)}-1
\right].
\label{eq:fex_hd}
\end{align}
Combining Eqs.~\eqref{eq:fid_hd} and~\eqref{eq:fex_hd}, the leading
high-dimensional free energy is
\begin{align}
-\beta f
&=
\frac{d\log d}{2}
\left[
\sum_{k=1}^m \binom{m}{k}g_k
-2m
\right]
\nonumber\\
&\quad
+
d\log d\,
\frac{\widetilde{\varphi}}{2}
\sum_{k=1}^m \binom{m}{k}g_k
\sum_{\ell=1}^k
\sum_{s=0}^{m-k}
\binom{k}{\ell}\binom{m-k}{s}
g_{\ell+s}
\int_{-\infty}^0 dh\,e^h
\left[
e^{-\ell\tilde{\beta}\tilde v(h)}-1
\right].
\label{eq:free_energy_hd}
\end{align}
In the zero-temperature limit, $\tilde{\beta}\to\infty$, this expression
reduces to the corresponding hard-sphere free energy.  The ideal-glass
transition density predicted by the conventional molecular ansatz
corresponds to $\widetilde{\varphi}=1$ at leading order~\cite{PZ2010}.
Interestingly, the free energy of hard spheres agrees with that of the
hard-core model on random graphs in the large degree limit, see
Appendix~\ref{sec:random_graph_hard_core}.

\subsection{Binomial ansatz and configurational entropy}

To obtain explicit results, we now restrict the molecular-size
distribution to the binomial form
\begin{align}
 g_k = x^{k-1}(1-x)^{m-k},
 \qquad 0\leq x\leq 1 .
\label{eq:binomial_ansatz}
\end{align}
This ansatz satisfies the normalization condition~\eqref{eq:normg}. The
limits $x=0$ and $x=1$ correspond, respectively, to the unbound liquid
ansatz $g_k=\delta_{k1}$ and the conventional full-replica molecular
ansatz $g_k=\delta_{km}$.

The parameter $x$ has a direct interpretation
as an inter-replica overlap. To see this, define
\begin{align}
Q_{ab}
\equiv
\frac{1}{N}
\sum_{i=1}^{N_{\rm mol}}
\left\langle
\mu_i^a\mu_i^b
\theta\left(\sigma-|x_i^a-x_i^b|\right)
\right\rangle,
\qquad a\neq b,
\end{align}
where $\sigma$ is a small microscopic length. In the high-dimensional
regime considered above, the cage size vanishes at leading order, and
the Heaviside factor is equal to one when replicas $a$ and $b$ belong to
the same molecule. Thus,
\begin{equation}
Q_{ab}
=
\frac{1}{N}
\sum_{i=1}^{N_{\rm mol}}
\left\langle
\mu_i^a\mu_i^b
\right\rangle
=
\sum_{\omu}g(\omu)\mu^a\mu^b
=
\sum_{k=2}^m
\binom{m-2}{k-2}g_k
=
x,
\end{equation}
suggesting that $x$ is the overlap between different replicas within the
binomial ansatz. It is also useful to define the replica-pair mismatch
fraction
\begin{align}
c
\equiv
\frac{1}{N}
\sum_{i=1}^{N_{\rm mol}}
\left\langle
\mu_i^a(1-\mu_i^b)
\right\rangle
=
1-x .
\label{eq:mismatch_fraction}
\end{align}
This quantity measures the fraction of particles in replica $a$ that
are not bound to their counterparts in replica $b$.

Using Eq.~\eqref{eq:binomial_ansatz}, the ideal contribution becomes
\begin{align}
-\beta f^{\rm id}
=
\frac{d\log d}{2}
\left[
\frac{1-(1-x)^m}{x}
-2m
\right],
\end{align}
and the excess contribution becomes
\begin{align}
-\beta f^{\rm ex}
=
\frac{d\log d}{2}
\widetilde{\varphi}
\int_{-\infty}^{0}dh\,e^h
\frac{
\left[
1+
\left(e^{-\tilde{\beta}\tilde v(h)}-1\right)x^2
\right]^m
-1
}{x^2}.
\end{align}
We now compute the configurational entropy by analytically continuing
the replicated free energy to real $m$. Let
\begin{align}
s(m,x)
\equiv
\frac{-\beta f(m,x)}{d\log d/2}.
\end{align}
The rescaled configurational entropy is obtained from the Monasson's
method~\cite{monasson1995}
\begin{align}
\tilde{\Sigma}(x)
\equiv
\frac{\Sigma(x)}{d\log d/2}
=
\left[
s(m,x)-m\partial_m s(m,x)
\right]_{m=1},
\label{eq:monasson_relation}
\end{align}
where the derivative is taken at fixed $x$. This gives
\begin{align}
\tilde{\Sigma}(x)
=
A(x)
-
\widetilde{\varphi}
\int_{-\infty}^0dh\,e^h\,
B(h)A\left(B(h)x^2\right),
\label{eq:sigma_compact}
\end{align}
where 
\begin{align}
A(x)
=
1+\frac{1-x}{x}\log(1-x),
\qquad
B(h)
=
1-e^{-\tilde{\beta}\tilde v(h)}.
\end{align}
This expression is the starting point for the analysis of the dynamical
and thermodynamic glass transitions in the next section.

\section{Results}
\label{sec:results}

\begin{figure}[t]
\begin{center}
\includegraphics[width=10cm]{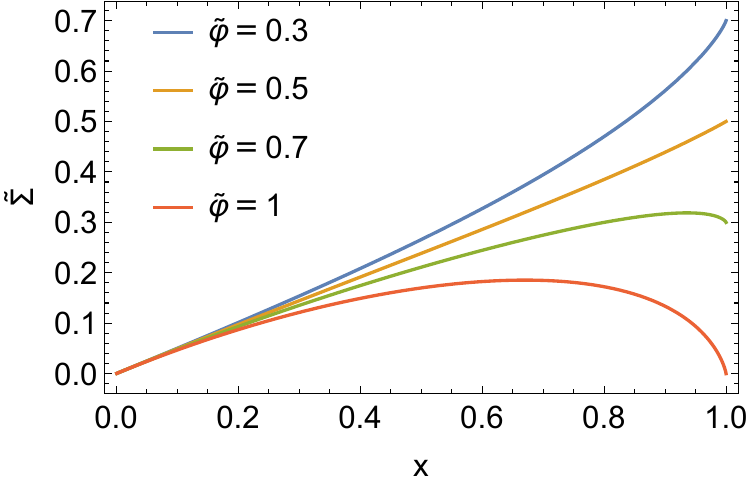}
\caption{
Configurational entropy of hard spheres. For $\tphi<1/2$,
$\tilde{\Sigma}(x)$ is monotonic and has its minimum at $x=0$. For
$\tphi>1/2$, a boundary minimum appears at $x=1$. This minimum becomes
the global minimum at $\tphi=1$.
}
\label{fig:sigma_hs}
\end{center}
\end{figure}

\subsection{Hard spheres}

We first consider the hard-sphere limit, $\tb\to\infty$. In this limit,
$B(h)\to 1$, and Eq.~\eqref{eq:sigma_compact} reduces to
\begin{align}
\tilde{\Sigma}(x)
=
A(x)-\tphi A(x^2).
\label{eq:sigma_hs}
\end{align}
Figure~\ref{fig:sigma_hs} shows the rescaled configurational entropy
$\tilde{\Sigma}(x)$. For $\tphi<1/2$, $\tilde{\Sigma}(x)$ is monotonic
and its minimum is located at $x=0$, corresponding to the liquid state.
For $\tphi>1/2$, a glassy minimum appears at the boundary $x=1$.

This boundary minimum is identified from the behavior near $x=1$.
For $1-x\ll 1$, Eq.~\eqref{eq:sigma_hs} gives
\begin{align}
\frac{\partial \tilde{\Sigma}}{\partial x}
\simeq
-(1-2\tphi)\log(1-x).
\end{align}
Since $\log(1-x)<0$, the endpoint $x=1$ becomes a local minimum when
$\tphi>1/2$. Thus the dynamical transition occurs at
\begin{align}
\tphi_d=\frac{1}{2},
\qquad
\varphi_d
=
\frac{2^{-d}d\log d}{2}.
\label{eq:phid_hs}
\end{align}
This revised hard-sphere dynamical transition reproduces the leading term $d\log d/2^{d+1}$
in the packing lower bound of Theorem~1.1 of Campos, Jenssen, Michelen,
and Sahasrabudhe~\cite{campos2023new}.   We
discuss this connection in more detail in
Appendix~\ref{224711_20Jun26}.
Whether a reversible
insertion--deletion process arrests at this density remains open (see Sec.~\ref{sec:dynint}).
A heuristic dynamical argument, given in Appendix~\ref{app:dyn_heuristic},
shows that the condition $\tphi>1/2$ for the stability of the endpoint
$x=1$ coincides, at leading order, with the condition under which a
particle removed by such a process is replaced within its own cage rather
than elsewhere.

The boundary glassy minimum becomes the global minimum at
$\tphi=1$. Therefore, the ideal-glass transition occurs at
\begin{align}
\tphi_K=1,
\qquad
\varphi_K
=
2^{-d}d\log d,
\label{eq:phik_hs}
\end{align}
which agrees with the leading high-dimensional result of the
conventional replica liquid theory~\cite{PZ2010}. Thus, for hard
spheres, molecular-size fluctuations shift the dynamical transition but
leave the leading ideal-glass transition density unchanged.

\subsection{Harmonic spheres at finite temperature}

\begin{figure}[t]
\begin{center}
\includegraphics[width=10cm]{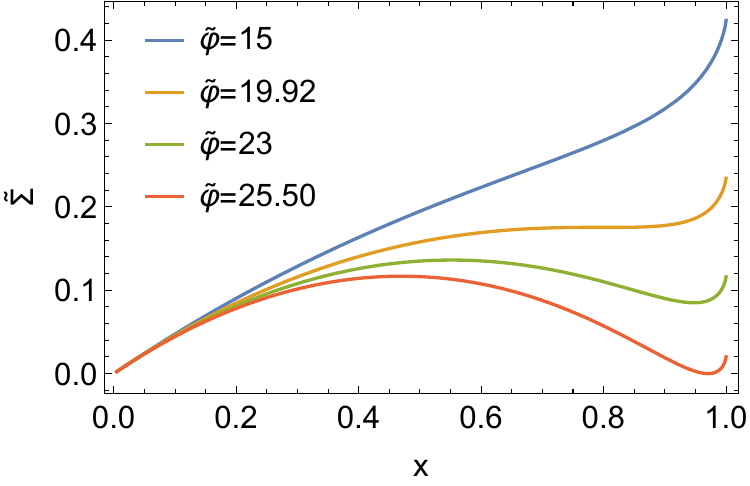}
\caption{
Configurational entropy of harmonic spheres for $\tb=0.2$. The glassy
minimum appears at $x_{\rm min}<1$. The dynamical and thermodynamic
transition points are $\tphi_d\simeq 19.92$ and $\tphi_K\simeq 25.50$,
respectively.
}
\label{fig:sigma_harmonic}
\end{center}
\end{figure}

We next consider harmonic spheres at finite temperature. In contrast to
the hard-sphere limit, the glassy minimum is no longer located at
$x=1$. As shown in Fig.~\ref{fig:sigma_harmonic}, the metastable glassy
solution appears at an interior point $x_{\rm min}<1$. This means that
the glassy state contains a finite replica-pair mismatch fraction at
finite temperature.

The position of the glassy minimum is determined by the stationarity
condition
\begin{align}
\frac{\partial \tilde{\Sigma}(x)}{\partial x}
=A'(x)-\tphi\int_{-\infty}^0 dh\, e^h\,
2x B(h)^2 A'\left(B(h)x^2\right)=0.
\label{eq:saddle_x}
\end{align}
Solving the stationarity condition for $\tphi$, one obtains
\begin{align}
\tphi(x)=
\frac{A'(x)}
{\int_{-\infty}^0 dh\, e^h\,2xB(h)^2 A'(B(h)x^2)}.
\end{align}
The dynamical transition is the spinodal at which a glassy stationary
point first appears, and is therefore given by
\begin{align}
\tphi_d=\min_{0<x<1}\tphi(x)
= \min_{0<x<1}\frac{A'(x)}
{\int_{-\infty}^0 dh\, e^h\,2xB(h)^2 A'(B(h)x^2)}.\label{eq:phid_harmonic}
\end{align}
The thermodynamic ideal-glass transition is obtained when the value of
$\tilde{\Sigma}(x)$ at the secondary minimum reaches zero. Equivalently,
it is the lowest density for which $\tilde{\Sigma}(x)=0$ has a solution
for $x>0$.  This gives
\begin{align}
\tphi_K
=
\min_{0<x<1}
\frac{
A(x)
}{
\int_{-\infty}^0 dh\, e^h\,
B(h)A\left(B(h)x^2\right)
}.
\label{eq:phik_harmonic}
\end{align}
Equations~\eqref{eq:phid_harmonic} and~\eqref{eq:phik_harmonic} allow us
to compute the dynamical and thermodynamic transition points very easily.

\begin{figure}[t]
\begin{center}
\includegraphics[width=10cm]{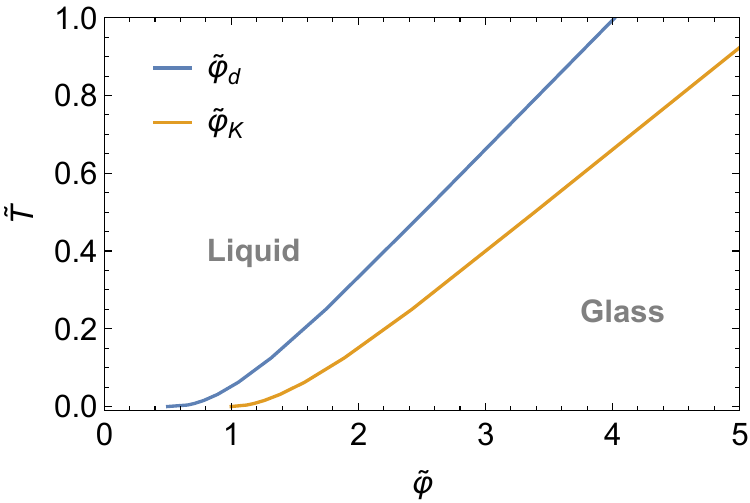} \caption{ Phase diagram
of high-dimensional harmonic spheres, 
where $\tilde{T}=\tilde{\beta}^{-1}$.  }
\label{fig:phase_harmonic}
\end{center}
\end{figure}

Figure~\ref{fig:phase_harmonic} shows the resulting phase diagram. The
region between $\tphi_d$ and $\tphi_K$ corresponds to a metastable
glassy phase with a finite replica-mismatch fraction. At $\tphi_K$, the
configurational entropy of this glassy minimum vanishes, $\tS(x_{\rm
min})=0$.

\subsection{Replica-mismatch fraction at low temperature}

We now analyze the low-temperature behavior of the replica-pair mismatch
fraction,
\begin{align}
c=1-x_{\rm min}.
\end{align}
At low temperature, $c$ is small. Setting $x=1-c$ with $c\ll 1$ and
using $A'(x)\simeq -\log c$, the stationarity condition
\eqref{eq:saddle_x} gives
\begin{align}
\log c
&\simeq
2\tphi
\int_{-\infty}^0 dh\, e^h
\log\left(e^{-\tb\tv(h)}+2c\right)
\nonumber\\
&\simeq
-2\tphi\tb
\int_{-\infty}^0 dh\, e^h
\min\left[
\frac{h^2}{2},
-\frac{\log c}{\tb}
\right].
\label{eq:lowT_c}
\end{align}
Introducing
\begin{align}
\lambda
=
-\frac{\log c}{\tb},
\label{eq:lambda_def}
\end{align}
Eq.~\eqref{eq:lowT_c} becomes
\begin{align}
\tphi
=
G(\lambda),
\qquad
G(\lambda)
\equiv
\frac{
\lambda
}{
2\int_0^\infty dt\, e^{-t}
\min\left[
\frac{t^2}{2},\lambda
\right]
}
=
\frac{
\lambda
}{
2\left[
1-(1+\sqrt{2\lambda})e^{-\sqrt{2\lambda}}
\right]
}.
\label{eq:G_lambda}
\end{align}
The function $G(\lambda)$ is defined for $\lambda\geq 0$ and satisfies
$G(0)=1/2$. Therefore, for $\tphi\geq 1/2$, Eq.~\eqref{eq:G_lambda}
admits a finite solution $\lambda(\tphi)$. The mismatch fraction then
obeys the Arrhenius form
\begin{align}
c
\sim
\exp\left[-\tb\lambda(\tphi)\right].
\label{eq:c_arrhenius}
\end{align}
Thus, the replica-mismatch fraction is exponentially suppressed at low
temperature. Nevertheless, the optimized glassy solution still has a
finite mismatch fraction at the thermodynamic glass transition point of
the harmonic sphere system.

\begin{figure}[t]
\begin{center}
\includegraphics[width=10cm]{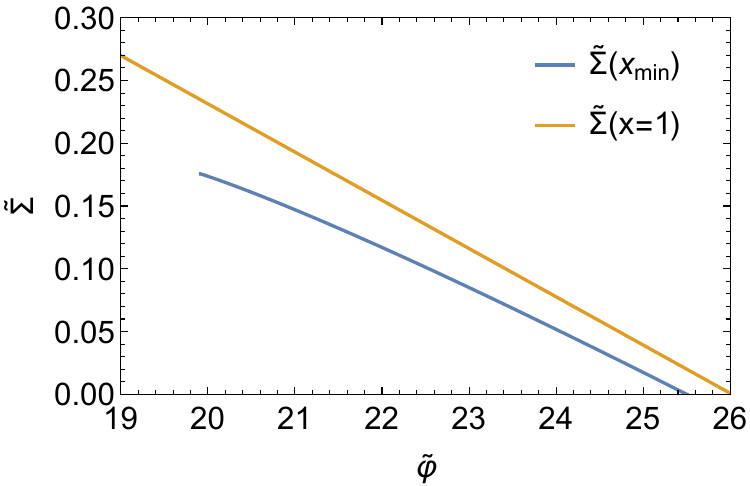} \caption{
Configurational entropy of harmonic spheres at $\tb=0.2$, evaluated at
$x_{\rm min}$ and $x=1$.
The conventional replica liquid theory, corresponding to $x=1$,
overestimates the configurational entropy.} \label{fig:two_sigma}
\end{center}
\end{figure}

\begin{figure}[t]
\begin{center}
\includegraphics[width=10cm]{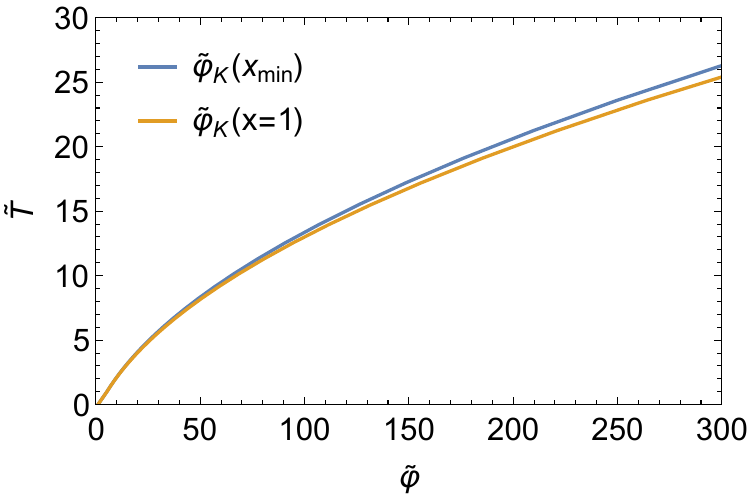} \caption{ Thermodynamic glass
transition point of harmonic spheres. The optimized result at $x_{\rm
min}$ is compared with the conventional replica liquid theory at $x=1$,
which overestimates the transition density, or equivalently
underestimates the transition temperature.} \label{fig:two_phik}
\end{center}
\end{figure}

\subsection{Comparison with the conventional theory}

Finally, we compare the results obtained including molecular
dissociation with the conventional replica liquid theory. The latter is
recovered by setting $x=1$, namely by imposing that every molecule
contains one particle from every replica. Figure~\ref{fig:two_sigma}
compares the configurational entropy evaluated at the optimized value
$x_{\rm min}$ with that evaluated at $x=1$. Since
\begin{align}
\tilde{\Sigma}(x_{\rm min})
<
\tilde{\Sigma}(x=1),
\end{align}
the conventional molecular ansatz overestimates the configurational
entropy of the glass.

This difference shifts the predicted thermodynamic transition point.
Figure~\ref{fig:two_phik} compares the transition point obtained from
the optimized ansatz with that obtained by imposing $x=1$. Since the
full-replica molecular ansatz overestimates the configurational entropy,
it also overestimates the transition packing fraction, or equivalently
underestimates the transition temperature.

This comparison also clarifies the physical meaning of conventional
estimates of the configurational entropy. Conventional numerical
estimates usually obtain the configurational entropy by subtracting the
vibrational entropy from the total
entropy~\cite{Frenkel1984,sciortino1999,Berthier2019}.  In the language
of replica liquid theory, this corresponds to the full-replica molecular
ansatz, in which all replicated particles remain bound around common
molecular centers and undergo only vibrational motions. As shown above,
this ansatz overestimates the configurational entropy and can therefore shift the ideal-glass
transition point. This interpretation is consistent with simulations of
glass forming systems, where localized excitations are present even inside the ideal-glass
phase~\cite{ozawa2018,jung2025numerical}.

\section{Summary and discussion}
\label{sec:summary}

\subsection{Summary}

In this work, we revisited the replica liquid theory of the glass
transition by relaxing the conventional full-replica molecular ansatz,
where each molecule contains one particle from every replica. We
generalized this construction by allowing each molecule to contain only
a subset of replicas. This introduces fluctuations in the molecular size
and provides a simple way to describe particle-level replica mismatches
within the replica liquid theory framework. We derived the free-energy
functional of this replicated molecular system including dissociation,
within the second-virial approximation and applied it to
high-dimensional hard and harmonic spheres. In the large-dimensional
limit, the theory simplifies considerably. Within the binomial ansatz,
the molecular-size distribution is controlled by a single parameter $x$,
which coincides with the overlap between different replicas.  For hard
spheres, molecular-size fluctuations destabilize the conventional glassy
metastable solution and shift the dynamical transition from the
conventional scale $O(2^{-d}d)$ to $O(2^{-d}d\log d)$. By contrast, the
leading thermodynamic transition density remains unchanged. For harmonic
spheres at finite temperature, the glassy metastable state appears at
$x<1$, indicating a finite mismatch fraction between replicas. This
mismatch fraction remains finite even at the thermodynamic glass
transition, thereby shifting the transition point from that predicted by
the conventional full-replica ansatz, even in the high-dimensional
limit.

\subsection{Dynamical interpretation}
\label{sec:dynint}

Our revised hard-sphere dynamical transition reproduces the leading term $d\log d/2^{d+1}$
in the packing lower bound of Theorem~1.1 of Campos, Jenssen, Michelen,
and Sahasrabudhe~\cite{campos2023new}, whose construction is summarized
in Appendix~\ref{224711_20Jun26}. Whether a reversible
insertion--deletion process arrests at this density remains open.  The
analogy with hard-core lattice gas problems on random graphs, discussed
in Appendix~\ref{sec:random_graph_hard_core}, suggests that the
relevant dynamics are those that do not conserve the number of
interacting particles.

More precisely, consider the Monte Carlo dynamics of $N$ particles, each
having position $x_i$ and carrying an additional binary degree of
freedom $\mu_i \in\{0,1\}$. When $\mu_i=0$, particle $i$ is a
non-interacting ideal gas particle; when $\mu_i =\mu_j=1$, then the
particles interact with potential $v(x_i-x_j)$. One can consider a
dynamics satisfying detailed balance, in which for instance the $x_i$
perform a Langevin dynamics with pair potential $\mu_i \mu_j v(x_i -
x_j)$, and from time to time one of the $\mu_i$ is flipped with the
Metropolis rule, adding a chemical potential that fixes the density of
$\mu=1$. The invariant measure of this system is that of a mixture.
Integrating out the $N-k$ inactive particles leaves a factor
$V^{N-k}/(N-k)!$ multiplying the usual weight of the $k$ active ones, so
that, even for an ideal gas, the number of active particles is
binomially distributed. This measure reduces to a grand-canonical one
for the liquid of particles with $\mu=1$ and pair potential $v(r)$ only
in the reservoir limit $N/k\to\infty$ at fixed effective activity
$zN/V$, where $z$ is the bare activity of the state $\mu=1$: in this
limit $V^{N-k}/(N-k)!=(V^N/N!)\,(N/V)^k\,e^{O(k^2/N)}$, so that the $k$
active particles carry the grand-canonical weight $(zN/V)^k/k!$ up to
corrections that vanish per particle.
Equivalently, one may consider genuine insertion and deletion moves, in
which a particle is inserted at a uniformly random position at rate
proportional to the activity and deleted at unit rate.  The long-time
limit of this dynamics can be described using
replicas~\cite{maimbourg2016,simpleglass} and leads precisely to a
replicated partition function of the form~\eqref{eq:XiGC}, after the
ideal-gas molecules with $|\omu|=0$ are once again integrated away.

We thus conjecture that the dynamical transition $\varphi_d \sim d \log d
/ 2^{d+1}$ given by the molecular ansatz including dissociation
corresponds to a class of dynamics that do not conserve the number of
interacting particles, such as the one described above in the reservoir
limit; we have not derived this arrest density from those dynamics.
A non-rigorous argument in favour of this conjecture is given in
Appendix~\ref{app:dyn_heuristic}: after a deletion, the vacated cage is
refilled faster than a particle is inserted elsewhere in its neighbourhood
if and only if $\tphi>1/2$, at leading order in high dimensions. The same
argument shows that, in the ghost-particle realization described above,
the inactive particles must lose memory of their position quickly
enough, so that a particle that has just been deactivated is not
privileged in refilling its own cage.
Instead, standard particle-number-conserving dynamics such as pure
Langevin become arrested at much lower density, $\varphi_d \simeq
4.8067\, d\, 2^{-d}$ in infinite dimension, as predicted by the
conventional molecular
ansatz~\cite{PZ2010,maimbourg2016,simpleglass}.  The fact that different dynamical
rules can give widely different dynamical arrest densities is not
surprising, and is well established e.g. in the case of the swap Monte
Carlo algorithm~\cite{NBC17}. The fact that replica theory, which is
based on thermodynamics, can capture these differences is also well
understood, because the replicated free energy is a Franz-Parisi
potential whose form depends on the choice of the molecular coupling,
different choices corresponding to different dynamical rules and hence
different dynamical order parameters~\cite{IZI17,ghimenti2026clever}.

If this conjecture holds, dynamics that do not conserve particle number
(such as grand-canonical Monte Carlo) would remain ergodic up to much
higher densities in high dimensions than particle-conserving dynamics
such as Langevin, Newtonian, or Monte Carlo dynamics.  A possible origin of the difference
lies in the peculiar properties of high-dimensional sphere packings, which feature a very diluted,
random-graph-like, network of particle contacts immersed in large void
regions. Strict particle conservation can lead to jamming in a
non-saturated packing with large enough voids, while the possibility to
remove and/or add particles can lead to a properly saturated packing
where no void can be occupied by an additional sphere.

\subsection{Localized excitations}

For harmonic spheres at finite temperature, the glassy solution is
located at $x<1$, rather than at the conventional value $x=1$. The
glassy state therefore contains a finite fraction of mismatch between
replicas, which follows the Arrhenius law. The conventional theory
restricted to $x=1$ overestimates the configurational entropy, and
therefore gives a finite configurational entropy even at the ideal-glass
transition. This interpretation is consistent with computer simulation
in randomly pinned
systems~\cite{cammarota2012ideal,ozawa2015equilibrium}, where the
configurational entropy remains finite even inside the ideal-glass
phase, due to the contributions of localized
excitations~\cite{ozawa2018}. More generally, the existence of localized
excitations within the glass is well known~\cite{Phillips1987} and the
present approach could be a first step toward including these
excitations within mean field theory.

\subsection{Perspectives}

Several limitations should be kept in mind. First, most explicit
calculations in the main text were performed within the binomial ansatz
for $g_k$. Appendix~\ref{141415_2Jul26} shows that the thermodynamic
transition is unchanged within a broader class of $g_k$, but a complete
analysis remains an open problem.  Second, the present theory is
static. In infinite-dimensional particle systems, the dynamical
transition obtained from the conventional replica calculation with $x=1$
is known to coincide with that obtained from Langevin
dynamics~\cite{maimbourg2016,szamel2017,simpleglass}. This agreement is
naturally understood from the fact that activated hopping and cage
escape are suppressed in the infinite-dimensional
limit~\cite{biroli2021interplay}. From this viewpoint, the present
formulation may be regarded as a static way of incorporating
hopping-like local rearrangements through replica mismatches. Note,
however, that interpreting the mismatch as hopping requires a trajectory
observable, since particles can exchange sites while the overlap of the
two density profiles remains high. Developing
a dynamical mean field theory that explicitly includes such processes
remains an important open problem, as discussed above; the heuristic
argument of Appendix~\ref{app:dyn_heuristic} only compares the leading
exponential behaviour of the relevant rates.  Such a dynamical
extension of the present theory may also clarify the relation between
mean field theory and the dynamical facilitation
scenario~\cite{chandler2010,keys2011}, where relaxation is controlled by
sparse localized excitations. The molecular dissociation mechanism
introduced here may be viewed as a static counterpart of such
excitations. A theory describing their creation and propagation would
provide a possible bridge between replica liquid theory and dynamical
facilitation.

\begin{acknowledgments}
We thank P.~Charbonneau, L.~Cugliandolo, G.~Folena, A.~Ikeda, M.~Jenssen, 
J.~Kurchan, P.~Morse, Y.~Nishikawa, F.~Ricci-Tersenghi, and G.~Szamel for
useful discussions.
This project has received JSPS KAKENHI Grant Numbers 23K13031 and
25H01401.

The initial submission was prepared with assistance from ChatGPT (OpenAI, GPT-5.6 Sol), which proposed the argument presented in Appendix B and helped refine the wording throughout the manuscript.
To address the suggestions of an anonymous referee, which we warmly thank, we used
the help of Claude (Anthropic, Opus 5.5); prompted by us, Claude derived the argument in Appendix D
and wrote the associated text.

\end{acknowledgments}

\appendix

\section{Heuristic summary of Campos {\it et al.}}
\label{224711_20Jun26}

To make the present paper self-contained, we briefly summarize the
recent lower bound on high-dimensional sphere packings obtained by
Campos, Jenssen, Michelen, and Sahasrabudhe (CJMS)~\cite{campos2023new}.
Their Theorem~1.1 guarantees a packing of density at least
$(1-o(1))\,d\log d/2^{d+1}$. It leaves the saturation density of
ordinary continuum random sequential addition (RSA) undetermined.
Saturation densities of continuum RSA have been estimated numerically
through $d=6$ by Torquato, Uche, and Stillinger~\cite{TUS06}, and through $d=8$ by Zhang and Torquato~\cite{ZhangTorquato2013};
unfortunately these numerical results are still too far from the asymptotic regime to draw conclusions on the scaling for $d\to\infty$.
Torquato considered a $d\log d\,2^{-d}$ term in Eq.~(94) of Ref.~\cite{Torquato2010}. A replica construction
that explicitly encodes the RSA insertion history was developed by
Jadrich, Lindquist, and Truskett~\cite{Jadrich2022}.

The purpose of this appendix is not to reproduce the rigorous proof of CJMS,
but to give a heuristic interpretation of the result in a form useful
for comparison with the replica liquid theory.
The key idea is to reduce the packing problem to an independent-set
problem on a random geometric graph. One samples $M$ points uniformly at
random in a volume $V$ and regards them as candidate centers of spheres
of diameter $D$. A graph is then constructed by connecting two points
whenever the corresponding spheres overlap. An independent set $\mathcal
I$ of this graph is a subset of vertices with no edges between
them. Therefore, it corresponds to a subset of candidate centers whose
spheres do not overlap, and hence defines a valid sphere packing.
In brief, CJMS first sample a Poisson set of candidate centers and prune
it, removing the points whose degree or codegree in the overlap graph is
too large. They then build an independent set in rounds, selecting
vertices in random batches and adding auxiliary edges
between rounds to keep the remaining graph approximately regular.

Let $|\mathcal I|$ be the size of the independent set, and define its
relative size by
\begin{align}
 \alpha = \frac{|\mathcal I|}{M}.
 \label{eq:CJMS_alpha}
\end{align}
The expected degree of the geometric graph is
\begin{align}
 K = \frac{M}{V} v_d(D),
 \label{eq:CJMS_K}
\end{align}
up to boundary corrections. The packing fraction associated with the
independent set is
\begin{align}
 \varphi
 =
 \frac{|\mathcal I|}{V}v_d\left(\frac{D}{2}\right).
\end{align}
Using $v_d(D)=2^d v_d(D/2)$, we obtain the relation
\begin{align}
 K = \frac{2^d\varphi}{\alpha}.
 \label{eq:CJMS_K_phi}
\end{align}

If the random geometric graph behaved as an Erd\H{o}s--R\'enyi random graph
with average degree $K$, a greedy algorithm, or a randomized greedy
procedure such as the R\"odl nibble method, would construct an
independent set of relative size
\begin{align}
 \alpha_{\rm alg}
 \simeq
 \frac{\log K}{K},
 \label{eq:CJMS_alpha_alg}
\end{align}
up to subleading corrections~\cite{Shearer1983,CojaOghlan2014}. By
contrast, the maximum independent set of an Erd\H{o}s--R\'enyi random
graph has the larger asymptotic size
\begin{align}
 \alpha_{\rm MIS}
 \simeq
 \frac{2\log K}{K}.
\end{align}
Thus, even for ordinary random graphs, there is a factor-of-two gap
between the size reached by known local constructive algorithms and the
maximum independent-set size.

In a random geometric graph, however, edges are not independent. Nearby
vertices induce strong local correlations among edges, and it is
therefore nontrivial whether the random-graph estimate
Eq.~\eqref{eq:CJMS_alpha_alg} remains attainable. The main achievement
of CJMS is to show that this algorithmic scaling does remain attainable
for high-dimensional geometric graphs in a regime where the average
degree grows as
\begin{align}
 K \simeq \Delta =
 \left(\frac{\sqrt{d}}{4\log d}\right)^{d},
 \label{eq:CJMS_K_scale}
\end{align}
where $\Delta$ is the degree parameter of Lemma~2.1 of
Ref.~\cite{campos2023new}. Combining
Eqs.~\eqref{eq:CJMS_K_phi}, \eqref{eq:CJMS_alpha_alg}, and
\eqref{eq:CJMS_K_scale}, we find, at the level of leading asymptotics,
\begin{align}
 \varphi_{\rm alg}
 =
 \frac{\alpha_{\rm alg}K}{2^d}
 \simeq
 \frac{\log K}{2^d}
 \simeq
 \frac{d\log d}{2^{d+1}}.
\end{align}
The leading term follows because $\log\Delta = \frac{d}{2}\log d
- d\log(4\log d) = \frac{d}{2}\log d\,[1+O(\log\log d/\log d)]$.
This is precisely the density scale of the dynamical transition
obtained in the present dissociation replica liquid theory.
Interestingly, at the same leading order, $\Delta$ coincides with the
ratio between the volume $v_d(D)$ excluded by a sphere and the volume of
its cage in the glass, i.e. the discretization of CJMS corresponds to
roughly one candidate center per cage (see
Appendix~\ref{app:dyn_heuristic}).

The factor-of-two gap between $\alpha_{\rm alg}$ and $\alpha_{\rm MIS}$
also provides a useful heuristic interpretation of the ideal-glass
transition. If a geometric analogue of the maximum-independent-set scale
$\alpha_{\rm MIS}\simeq 2\log K/K$ were attainable, the corresponding
packing fraction would be
\begin{align}
 \varphi_{\rm MIS}
 =
 \frac{\alpha_{\rm MIS}K}{2^d}
 \simeq
 \frac{2\log K}{2^d}
 \simeq
 \frac{d\log d}{2^d}.
\end{align}
This has the same leading scale as the ideal-glass transition density
$\varphi_K$ and the glass close packing density $\varphi_{GCP}$
predicted by the replica liquid theory~\cite{PZ06a,PZ2010}; CJMS already
noted this factor-of-two comparison in their Sec.~1.2.  This comparison
suggests that the difference between the CJMS lower bound and
$\varphi_K$ or $\varphi_{GCP}$ may be interpreted heuristically as an
algorithmic gap: the CJMS construction reaches the
independent-set scale $\log K/K$, whereas the thermodynamic threshold
corresponds to the larger scale $2\log K/K$.
We stress that $\varphi_K$ and $\varphi_{GCP}$ are replica predictions
for amorphous states obtained by adiabatic compression of the liquid state, not bounds on the densest packing: Klartag's lower
bound $c\,d^2 2^{-d}$ for lattice packings in every $d\geq
2$~\cite{Klartag2026}, with universal $c>0$, asymptotically exceeds both
glass scales. The CJMS result remains relevant to the present comparison
because it is obtained from a random construction.

\section{More general ansatz}
\label{141415_2Jul26}

In the main text, we analyzed the model within the binomial ansatz.
In this Appendix, we show that the resulting thermodynamic glass
transition is unchanged within a broader class of ans\"atze. Following
the treatment of the hard-core lattice gas on random
graphs~\cite{weigt2001}, we consider
\begin{align}
g(\omu)
=
\int dh\, P(h)\,
\frac{\exp\left(h\sum_{a=1}^m \mu^a\right)}
{(1+e^h)^m}.
\end{align}
Introducing
\begin{align}
x=\frac{e^h}{1+e^h},
\end{align}
this ansatz can be rewritten as
\begin{align}
g_k
=
\int_0^1 dx\, p(x)\, x^{k-1}(1-x)^{m-k},
\end{align}
where
\begin{align}
p(x)\equiv x P(h)\frac{dh}{dx}.
\end{align}
The normalization condition gives
\begin{align}
\int_0^1 dx\, p(x)=1.
\end{align}
Thus, $p(x)$ can be interpreted as the probability distribution of the
local occupation probability $x$. The set of all probability
distributions $p(x)$ is convex: if $p_1(x)$ and $p_2(x)$ are probability
distributions, then
$p_\lambda(x)=\lambda p_1(x)+(1-\lambda)p_2(x)$, $0\leq \lambda \leq 1$
is also a probability distribution. The binomial ansatz used in the
main text corresponds to an extreme point of this convex set,
\begin{align}
p(x)=\delta(x-x_0).
\end{align}

Within this dissociation ansatz, the ideal-gas contribution is
\begin{align}
-\beta f^{\rm id}
=
\frac{d}{2}\log d
\int_0^1 dx\, p(x)
\left[
\frac{1-(1-x)^m}{x}
-2m
\right].
\end{align}
The excess contribution is
\begin{align}
-\beta f^{\rm ex}
=
\frac{d\log d}{2}\tphi
\int_{-\infty}^0 dh\, e^h
\int_0^1 dx\,dy\, p(x)p(y)
\left[
\frac{\left\{1+\left(e^{-\tb \tv(h)}-1\right)xy\right\}^m-1}{xy}
\right].
\end{align}
Then, the configurational entropy is obtained as
\begin{align}
\tS[p]
=
\int_0^1 dx\, p(x) A(x)
-
\tphi
\int_{-\infty}^0 dh\, e^h B(h)
\int_0^1 dx\,dy\, p(x)p(y) A\!\left(B(h)xy\right),
\end{align}
where 
\begin{align}
A(x)
&=
1+\frac{(1-x)\log(1-x)}{x}
=
\sum_{n=1}^{\infty}\frac{x^n}{n(n+1)},
\\
B(h)
&=
1-e^{-\tb \tv(h)}.
\end{align}
For a repulsive interaction, $0\leq B(h)\leq 1$.

We now show that allowing a distribution $p(x)$ cannot lower the
configurational entropy below the result obtained within the binomial
ansatz. First, $\tS[p]$ is a concave functional of
\(p(x)\). Indeed, for
\begin{align}
p_\lambda(x)=\lambda p_1(x)+(1-\lambda)p_2(x),
\qquad
0\leq \lambda\leq 1,
\end{align}
one finds
\begin{align}
\tS[p_\lambda]
-\lambda \tS[p_1]
-(1-\lambda)\tS[p_2]
=
\tphi
\int_{-\infty}^0 dh\, e^h
\sum_{n=1}^{\infty}
\frac{B(h)^{n+1}}{n(n+1)}
\lambda(1-\lambda)
\left(\ave{x^n}_1-\ave{x^n}_2\right)^2
\geq 0,
\end{align}
where
\begin{align}
\ave{x^n}_i
=
\int_0^1 dx\, p_i(x)x^n
\qquad
(i=1,2).
\end{align}
The concavity of $\tS[p]$ suggests, in accordance with Bauer's maximum
principle, that the global minimum should be realized at an extreme
point of the convex set of probability measures, $p(x)=\delta(x-x_{\rm
min})$~\cite{aliprantis2006}.  Below we verify this conclusion directly
for the present functional.
To this end, we define 
\begin{align}
\tS(x)
=
A(x)
-
\tphi
\int_{-\infty}^0 dh\, e^h B(h)
A\!\left(B(h)x^2\right),
\end{align}
which is precisely the configurational entropy obtained from the
binomial ansatz. Then
\begin{align}
&\tS[p]-\int_0^1 dx\, p(x)\tS(x)\new 
&=
\tphi
\int_{-\infty}^0 dh\, e^h B(h)
\left[
\int_0^1 dx\, p(x) A\!\left(B(h)x^2\right)
-
\int_0^1 dx\,dy\, p(x)p(y)
A\!\left(B(h)xy\right)
\right]\new
&=
\tphi
\int_{-\infty}^0 dh\, e^h
\sum_{n=1}^{\infty}
\frac{B(h)^{n+1}}{n(n+1)}
\left[
\ave{x^{2n}}-\ave{x^n}^2\right]
 \geq 0.
\end{align}
The last inequality follows from the positivity of the variance,
\begin{align}
\ave{x^{2n}}-\ave{x^n}^2\geq 0.
\end{align}
Therefore,
\begin{align}
\tS[p]
\geq
\int_0^1 dx\, p(x)\tS(x)
\geq
\min_{0\leq x\leq 1}\tS(x).
\end{align}
Hence the global minimum of the configurational entropy within the
dissociation ansatz is identical to that obtained within the binomial
ansatz.  Consequently, the thermodynamic glass transition discussed in
the main text is unchanged by allowing fluctuations of the local
occupation probability $x$.

We finally comment on the dynamical transition point. Since $\tS[p]$ is
concave in $p$, a metastable minimum associated with the dynamical
transition cannot be realized by any non-extreme distribution. Indeed,
$\tS[p_\lambda]$ can be considered as a one-dimensional concave function
of $\lambda$, which cannot have a local minimum at an interior point.

\section{Hard-core model on random graphs in the large-degree limit}
\label{sec:random_graph_hard_core}

In this appendix, we discuss the hard-core model on an
Erd\H{o}s--R\'enyi random graph with mean degree $K$. The hard-core
model on a graph is equivalent to the problem of independent sets:
occupied vertices correspond to vertices included in the independent
set, and the hard-core constraint forbids two adjacent vertices from
being occupied simultaneously. Thus, each allowed configuration of the
hard-core model is an independent set of the graph. The purpose of this
appendix is to show that, in the large-$K$ limit, the replica free
energy of this problem has the same structure as the one obtained for
high-dimensional hard spheres in the main text.

\subsection{Settings}

We consider a graph $G=(V,E)$ with $M$ vertices. We introduce an
occupation variable $\mu_i\in\{0,1\}$ on each vertex. If $\mu_i=1$,
vertex $i$ is occupied and belongs to the independent set; if
$\mu_i=0$, it is empty. The hard-core constraint requires that no edge
connects two occupied vertices. Therefore, the partition function
counting independent sets is
\begin{align}
 Z =
 \sum_{\{\mu_i=0,1\}}
 \prod_{(ij)\in E}(1-\mu_i\mu_j).
\end{align}
The factor $(1-\mu_i\mu_j)$ vanishes when both endpoints of an edge are
occupied, and is otherwise equal to one. Hence only independent sets
contribute to $Z$.

We are interested in the entropy of independent sets at fixed density
\begin{align}
 \alpha
 =
 \frac{1}{M}\sum_{i=1}^M \mu_i .
\end{align}
In the replica calculation below, this density is imposed as a constraint
on each replica.

\subsection{Replica free energy}

The disorder-averaged free energy is computed from
\begin{align}
 -\beta f
 =
 \frac{1}{M}\overline{\log Z}
 =
 \lim_{n\to 0}
 \frac{1}{nM}\log \overline{Z^n}.
\end{align}
The replicated partition function is
\begin{align}
 Z^n
 =
 \sum_{\{\mu_i^a=0,1\}}
 \prod_{(ij)\in E}
 \prod_{a=1}^n
 (1-\mu_i^a\mu_j^a).
\end{align}
For an Erd\H{o}s--R\'enyi graph in which each edge is present
independently with probability $K/M$, the average over graphs
gives~\cite{weigt2001}
\begin{align}
\overline{Z^n}
&=
\sum_{\{\mu_i^a=0,1\}}
\prod_{i<j}
\left[
1-\frac{K}{M}
+
\frac{K}{M}
\prod_{a=1}^n
(1-\mu_i^a\mu_j^a)
\right]
\nonumber\\
&=
\sum_{\{\mu_i^a=0,1\}}
\exp\left[
-\frac{KM}{2}
+
\frac{K}{2M}
\sum_{ij}
\prod_{a=1}^n
(1-\mu_i^a\mu_j^a)
+O(1)
\right].
\end{align}

We introduce the replicated order parameter
\begin{align}
 \rho(\omu)
 =
 \frac{1}{M}
 \sum_{i=1}^M
 \prod_{a=1}^n
 \delta_{\mu_i^a,\mu^a},
 \qquad
 \omu=\{\mu^1,\cdots,\mu^n\}.
\end{align}
It satisfies the normalization condition
\begin{align}
 \sum_{\omu}\rho(\omu)=1.
\end{align}
The fixed-density constraint in each replica is
\begin{align}
 \alpha
 =
 \sum_{\omu}\rho(\omu)\mu^a,
 \qquad
 a=1,\cdots,n.
\end{align}

Using the standard saddle-point representation of the order parameter,
one obtains
\begin{align}
\overline{Z^n}
=
\int D\rho(\omu)
\exp\left[-M\beta f_n[\rho]\right],
\end{align}
with
\begin{align}
-\beta f_n[\rho]
=
-\sum_{\omu}\rho(\omu)\log\rho(\omu)
-\frac{K}{2}
+
\frac{K}{2}
\sum_{\omu,\onu}
\rho(\omu)\rho(\onu)
\prod_{a=1}^n
(1-\mu^a\nu^a).
\label{eq:rg_replica_free_energy}
\end{align}
The free energy should be extremized under the normalization and
fixed-density constraints.

\subsection{Factorized 1RSB ansatz}

We now consider a one-step replica-symmetry-broken ansatz. The $n$
replicas are divided into $n/m$ blocks of size $m$, and the order
parameter is assumed to factorize over the blocks:
\begin{align}
 \rho(\mu^1,\cdots,\mu^n)
 =
 \prod_{r=1}^{n/m}
 \rho_1(\mu^{(r-1)m+1},\cdots,\mu^{rm}),
\end{align}
which is tantamount to assuming that $m$ replicas within the block belong to the same glassy metastable state, and that replicas from different blocks are uncorrelated.
Substituting this ansatz into Eq.~\eqref{eq:rg_replica_free_energy},
we obtain
\begin{align}
-\beta f_n
&=
-\frac{n}{m}
\sum_{\omu}
\rho_1(\omu)\log\rho_1(\omu)
\nonumber\\
&\quad
+\frac{K}{2}
\left[
\left(
\sum_{\omu,\onu}
\rho_1(\omu)\rho_1(\onu)
\prod_{a=1}^m
(1-\mu^a\nu^a)
\right)^{n/m}
-1
\right],
\end{align}
where now $\omu=\{\mu^1,\cdots,\mu^m\}$ is a block variable. Taking the
limit $n\to 0$, we find
\begin{align}
-\beta f
=
\lim_{n\to 0}\frac{-\beta f_n}{n}
=
-\frac{1}{m}
\sum_{\omu}
\rho_1(\omu)\log\rho_1(\omu)
+
\frac{K}{2m}
\log
\left[
\sum_{\omu,\onu}
\rho_1(\omu)\rho_1(\onu)
\prod_{a=1}^m
(1-\mu^a\nu^a)
\right].
\label{eq:rg_1rsb_free_energy}
\end{align}

\subsection{Small-$\alpha$ expansion}

In the large-$K$ limit, the density of large independent sets scales as
$\alpha=O(\log K/K)$, and is therefore small. We expand
Eq.~\eqref{eq:rg_1rsb_free_energy} for $\alpha\ll 1$.

Let $\overline{0}$ denote the empty block, $\mu^a=0$ for all $a$. Since
the density is small, $\rho_1(\overline{0})=O(1)$, while
$\rho_1(\omu)=O(\alpha)$ for $\omu\neq\overline{0}$. We write
\begin{align}
\rho_1(\omu)
=
\delta_{\omu,\overline{0}}\rho_1(\overline{0})
+
(1-\delta_{\omu,\overline{0}})\alpha g(\omu),
\end{align}
where $g(\omu)=\rho_1(\omu)/\alpha$ for $\omu\neq\overline{0}$. The
normalization gives
\begin{align}
 \rho_1(\overline{0})
 =
 1-\alpha\sum_{|\omu|>0}g(\omu).
\end{align}
The density constraint becomes
\begin{align}
 \sum_{|\omu|>0}g(\omu)\mu^a=1,
 \qquad
 a=1,\cdots,m.
\label{eq:rg_g_constraint}
\end{align}
Note that removing the $\overline{0}$ in this way precisely correspond to the idea of integrating away ideal gas particles discussed
in Sec.~\ref{sec:dynint}.
Assuming replica symmetry within each block, $g(\omu)$ depends only on
$|\omu|=\sum_{a=1}^m\mu^a$:
\begin{align}
 g(\omu)=g_{|\omu|}.
\end{align}
Then Eq.~\eqref{eq:rg_g_constraint} reduces to
\begin{align}
 \sum_{k=1}^m
 \binom{m-1}{k-1}g_k
 =
 1.
\label{eq:rg_gk_constraint}
\end{align}

The entropic term has the leading small-$\alpha$ behavior
\begin{align}
 \sum_{\omu}\rho_1(\omu)\log\rho_1(\omu)
 =
 \alpha\log\alpha
 \sum_{|\omu|>0}g(\omu)
 +O(\alpha).
\end{align}
The interaction term is
\begin{align}
&\log
\left[
\sum_{\omu,\onu}
\rho_1(\omu)\rho_1(\onu)
\prod_{a=1}^m
(1-\mu^a\nu^a)
\right]
\nonumber\\
&\qquad
=
\alpha^2
\sum_{|\omu|>0}
\sum_{|\onu|>0}
g(\omu)g(\onu)
\left[
\prod_{a=1}^m(1-\mu^a\nu^a)-1
\right]
+O(\alpha^3).
\end{align}
Therefore, to leading order,
\begin{align}
-m\beta f
&=
-\alpha\log\alpha
\sum_{|\omu|>0}g(\omu)
+
\frac{K\alpha^2}{2}
\sum_{|\omu|>0}
\sum_{|\onu|>0}
g(\omu)g(\onu)
\left[
\prod_{a=1}^m(1-\mu^a\nu^a)-1
\right]
\nonumber\\
&=
-\alpha\log\alpha
\sum_{k=1}^m
\binom{m}{k}g_k
-
\frac{K\alpha^2}{2}
\sum_{k=1}^m
\binom{m}{k}g_k
\sum_{\ell=1}^{k}
\sum_{s=0}^{m-k}
\binom{k}{\ell}
\binom{m-k}{s}
g_{\ell+s}.
\label{eq:rg_small_alpha_free_energy}
\end{align}
This can be identified with the free-energy calculated by the 
Monasson method~\cite{monasson1995}, since we assumed that the $m$ replicas are in the same glassy metastable state. Equation~(\ref{eq:rg_small_alpha_free_energy}) has the same functional form as the high-dimensional
hard-sphere free energy in the main text, Eq.~\eqref{eq:free_energy_hd},
in the hard-sphere limit $\tb\to\infty$, up to terms linear in $m$ that
do not affect the configurational entropy.
Note that to obtain a non-trivial results the two terms need to be of the same order, which forces
$\alpha \propto \log K/K$ in the independent set problem, and $\varphi \propto d \log d/2^{d}$ in the hard-sphere model. If these scalings are not
respected, the identification does not hold.

\subsection{Binomial ansatz}

As in the particle problem, we now impose the binomial ansatz
\begin{align}
 g_k
 =
 x^{k-1}(1-x)^{m-k}.
\end{align}
This ansatz satisfies Eq.~\eqref{eq:rg_gk_constraint}. Substituting it
into Eq.~\eqref{eq:rg_small_alpha_free_energy}, we obtain
\begin{align}
-m\beta f
=
-\alpha\log\alpha\,
\frac{1-(1-x)^m}{x}
-
\frac{K\alpha^2}{2}
\frac{1-(1-x^2)^m}{x^2}.
\label{eq:rg_binomial_free_energy}
\end{align}

The configurational entropy is obtained by the same Monasson derivative
as in the main text. Introducing the rescaled entropy
\begin{align}
\tilde{\Sigma}(x)
\equiv
\frac{\Sigma(x)}{-\alpha\log\alpha},
\end{align}
we find
\begin{align}
\tilde{\Sigma}(x)
=
A(x)
-
\widetilde{\varphi}^{\rm eff} A(x^2),
\label{eq:rg_sigma}
\end{align}
where
\begin{align}
 A(x)
 =
 1+\frac{1-x}{x}\log(1-x),
 \qquad
 \widetilde{\varphi}^{\rm eff}
 =
 -\frac{K\alpha}{2\log\alpha}.
\end{align}
Equation~\eqref{eq:rg_sigma} is identical to the hard-sphere
configurational entropy in the large-dimensional limit. Therefore, the
dynamical and thermodynamic transition points are
\begin{align}
 \widetilde{\varphi}^{\rm eff}_d=\frac{1}{2},
 \qquad
 \widetilde{\varphi}^{\rm eff}_K=1.
\end{align}
Solving these equations for large $K$ gives
\begin{align}
 \alpha_d
 =
 \frac{\log K}{K}
 +
 O\left(\frac{\log\log K}{K}\right),
 \qquad
 \alpha_K
 =
 \frac{2\log K}{K}
 +
 O\left(\frac{\log\log K}{K}\right).
\end{align}
The second result agrees with the well-known leading asymptotic density
of maximum independent sets on large random graphs, while the first
gives the corresponding dynamical, or algorithmic, scale. These results
are consistent with rigorous results and cavity-method analyses of the
hard-core model on random graphs~\cite{CojaOghlan2014,barbier2013}.
A simple dynamical interpretation of $\alpha_d$, in terms of the
competition between the refilling of a vacated vertex and the
occupation of one of its neighbours, is given in
Appendix~\ref{app:dyn_heuristic}.
See Ref.~\cite{angelini2019monte} for a comparison of different grand-canonical algorithms for the maximum independent set problem.

\section{Heuristic dynamical argument for the grandcanonical arrest}
\label{app:dyn_heuristic}

In this appendix we give a non-rigorous argument supporting the
conjecture of Sec.~\ref{sec:dynint}, namely that the dynamical
transition $\tphi_d=1/2$ of Eq.~\eqref{eq:phid_hs} is the density at
which a reversible insertion--deletion dynamics of hard spheres arrests.
The argument only compares the leading exponential behaviour, at order
$e^{O(d\log d)}$, of the rates of a few elementary processes.

\subsection{Insertion--deletion dynamics}

We consider hard spheres of diameter $D=1$ evolving with the following
dynamics: each particle is deleted with rate $1$, and a particle is
inserted with rate $z$ per unit volume at any point whose distance from
all particles is larger than $D$. This dynamics satisfies detailed
balance with respect to the grandcanonical measure with activity $z$.
In high dimensions and at the density scale
$\varphi=\tphi\,2^{-d}d\log d$, the second virial approximation is
exact~\cite{PZ2010}, and the probability that a random point can accept a new
particle is $e^{-2^d\varphi}$, so that
\begin{align}
\rho = z\,e^{-2^d\varphi},
\qquad
2^d\varphi = \rho\, v_d(D) = \tphi\, d\log d .
\label{eq:app_rho_z}
\end{align}
We also recall that, with $D=1$,
\begin{align}
\log\rho = \frac{d}{2}\log d\,[1+o(1)],
\qquad
\log v_d(D) = -\frac{d}{2}\log d\,[1+o(1)],
\end{align}
which follows from $\varphi=\rho\, v_d(D/2)$ and
$v_d(r)=\pi^{d/2}r^d/\Gamma(d/2+1)$.
If particles also move, for instance with Langevin dynamics as in the
ghost-particle realization of Sec.~\ref{sec:dynint}, then at
$\tphi=O(1)$ the density is much larger than the dynamical transition
$\varphi_d\simeq 4.8067\,d\,2^{-d}$ of particle-conserving dynamics, so
each interacting particle only rattles in its cage and the argument below is
unchanged at leading order.

\subsection{Return versus escape}

Consider a particle centered in $X_0$ and suppose that it is deleted.
The newly available space is the set of points that are within distance
$D$ of $X_0$ and farther than $D$ from all other particles. It consists
of two parts, which lead to two competing processes.

\emph{Return.} A new particle may be inserted in the cage left by the
deleted one, i.e. in the free region surrounding $X_0$. Its volume
$v_c$ is controlled by the cage size $A$ introduced in
Eq.~\eqref{eq:gaussian_ansatz}: $\log v_c\simeq\frac{d}{2}\log A$, and
since $d\log A=-2d\log d\,[1+o(1)]$ at this density scale, we have
$\log v_c=-d\log d\,[1+o(1)]$. The rate of this process is
\begin{align}
\Gamma_{\rm ret}=z\,v_c ,
\qquad
\log\Gamma_{\rm ret}
=\log\rho+2^d\varphi+\log v_c
=\left(\tphi-\frac12\right)d\log d\,[1+o(1)] .
\label{eq:app_Gret}
\end{align}

\emph{Escape.} A new particle may instead be inserted at a point $y$
that was excluded only by the deleted particle, at a distance of order
$D$ from $X_0$. In high dimensions the exclusion spheres of $X_0$ and
of a typical such point have negligible overlap, so the probability
that $y$ is not excluded by other particles is $e^{-2^d\varphi}$, and
the corresponding volume is $V_F\simeq v_d(D)\,e^{-2^d\varphi}$. Using
Eq.~\eqref{eq:app_rho_z}, the rate of this process is
\begin{align}
\Gamma_{\rm esc}=z\,V_F=\rho\, v_d(D)=2^d\varphi=\tphi\, d\log d .
\label{eq:app_Gesc}
\end{align}
An escape replaces the deleted particle by one that is not in its cage,
and therefore reduces the overlap between the density profiles at
different times, while a return preserves it.

Combining Eqs.~\eqref{eq:app_Gret} and~\eqref{eq:app_Gesc},
\begin{align}
\frac{\Gamma_{\rm ret}}{\Gamma_{\rm esc}}
=\exp\left[\frac{d\log d}{2}\,(2\tphi-1)\,[1+o(1)]\right].
\label{eq:app_ratio}
\end{align}
For $\tphi<1/2$, a deleted particle is replaced by an escape with
probability approaching one, and the overlap decays on the time scale of
deletions, which is of order one: the dynamics does not arrest. For
$\tphi>1/2$, the cage is refilled with probability
$1-O(\Gamma_{\rm esc}/\Gamma_{\rm ret})$. Moreover the refilling time
$1/\Gamma_{\rm ret}$ is much shorter than the time scale of all other
events, which justifies treating the rest of the system as frozen during
the refilling. The overlap then remains equal to one at leading order,
and decays only on a time scale of order
$\Gamma_{\rm ret}/\Gamma_{\rm esc}$, which diverges as $d\to\infty$:
the dynamics is arrested. The two regimes are separated by $\tphi=1/2$,
in agreement with Eq.~\eqref{eq:phid_hs}. In addition, the overlap
remains close to one in the arrested phase, as the boundary minimum
$x=1$ found in Sec.~\ref{sec:results}. 

\subsection{Random graphs}

The same argument applies to the hard-core model on random graphs with
average degree $K$ discussed in Appendix~\ref{sec:random_graph_hard_core},
with a Glauber dynamics in which an occupied vertex is emptied with rate
$1$ and an empty vertex with no occupied neighbour is occupied with rate
$\lambda$. When vertex $0$ is emptied, it can be occupied again with
rate $\lambda$ (return), or one of the $F$ neighbours whose only
occupied neighbour was $0$ can be occupied with total rate $\lambda F$
(escape). At density $\alpha$, a vertex has no occupied neighbour with
probability $e^{-K\alpha}$, hence $F\simeq K e^{-K\alpha}$, and
\begin{align}
\frac{\Gamma_{\rm ret}}{\Gamma_{\rm esc}}=\frac{e^{K\alpha}}{K} ,
\end{align}
which exceeds one if and only if $\alpha>\log K/K$. This is the
dynamical density $\alpha_d$ found in
Appendix~\ref{sec:random_graph_hard_core}. On random graphs, the
existence of an algorithmic barrier at this density is supported by
rigorous results: Coja-Oghlan and Efthymiou proved that the space of
independent sets of density larger than $(1+\varepsilon)\log K/K$ is
shattered, and that the Metropolis process needs an exponentially long
time to sample them~\cite{CojaOghlan2014}.

Comparing the two cases suggests the dictionary
$K\leftrightarrow v_d(D)/v_c$, i.e. one candidate vertex per cage.
Indeed, $\log[v_d(D)/v_c]=\frac{d}{2}\log d\,[1+o(1)]$, which coincides
at leading order with $\log K$ in the regime considered by CJMS,
Eq.~\eqref{eq:CJMS_K_scale}.

\subsection{Limitations}

The argument above is not a derivation of the dynamical arrest. It
only compares the leading exponential behaviour of the rates, and it
neglects corrections of order $e^{O(d)}$, which include the prefactor of
the cage size, the fluctuations of the cage volume, and the
cooperative effects associated with escapes, since a particle inserted
outside the vacated cage can itself block other cages. Therefore, it
cannot locate the transition within a window
$\tphi-1/2=O(1/\log d)$, nor describe the form of the relaxation close
to the transition. Moreover, the rates above are rates in physical
time. If insertions are attempted at uniformly random positions, only a
fraction $e^{-2^d\varphi}$ of the attempts is accepted, and the
computational cost of the dynamics is a separate question that we do not
address.

Finally, the argument assumes that a deleted particle has no privileged
role in the refilling of its own cage. This holds for the
insertion--deletion dynamics defined above, in which particles are
inserted at random positions. In the ghost-particle realization of
Sec.~\ref{sec:dynint}, a deactivated particle remains in its cage as an
ideal-gas particle and can be reactivated there, which provides an
additional return channel. This channel is negligible if the inactive
particles leave the cage quickly compared with their reactivation, or
if their density is high enough that the cage contains many of them,
which requires $N/k\gtrsim 1/(\rho\, v_c)$, i.e. a ratio between
inactive and active particles growing at least as
$e^{\frac{d}{2}\log d}$. Otherwise, the additional return channel
favours the refilling of the cage and could lead to arrest at a lower
density. Note that the replica calculation is consistent with the
former case, because the coordinates of the inactive replicas of a
molecule are integrated over the whole volume, i.e. inactive particles
are assumed to have lost memory of their position.

\bibliography{reference.bib,HSmerge.bib}

\end{document}